\documentclass[]{aastex631}

\graphicspath{{./}{figures/}}
\submitjournal{ApJL}

\shorttitle{Solar Modulation of helium nuclei with PAMELA}
\shortauthors{N. Marcelli et al.}

\begin{document}

\title{Helium fluxes measured by the PAMELA experiment from the minimum to the maximum solar activity for solar cycle 24}

\correspondingauthor{R. Munini}
\email{riccardo.munini@ts.infn.it}

\author{N. Marcelli} 
\affiliation{University of Rome ``Tor Vergata'', Department of Physics, I-00133 Rome, Italy}
\affiliation{INFN, Sezione di Rome ``Tor Vergata'', I-00133 Rome, Italy} 

\author{M. Boezio}
\affiliation{INFN, Sezione di Trieste, I-34149 Trieste, Italy} 
\affiliation{IFPU, I-34014 Trieste, Italy} 

\author{A. Lenni}
\affiliation{INFN, Sezione di Trieste, I-34149 Trieste, Italy} 
\affiliation{IFPU, I-34014 Trieste, Italy} 
\affiliation{University of Trieste, Department of Physics, I-34100 Trieste, Italy}

\author{W. Menn}
\affiliation{Universitat Siegen, Department of Physics, D-57068 Siegen, Germany}

\author{R. Munini} 
\affiliation{INFN, Sezione di Trieste, I-34149 Trieste, Italy} 
\affiliation{IFPU, I-34014 Trieste, Italy}

\author{O. P. M. Aslam}, 
\affiliation{North-West University, Centre for Space Research, 2520 Potchefstroom, South Africa}

\author{D. Bisschoff} 
\affiliation{North-West University, Centre for Space Research, 2520 Potchefstroom, South Africa}

\author{M. D. Ngobeni} 
\affiliation{North-West University, Centre for Space Research, 2520 Potchefstroom, South Africa}

\author{M. S. Potgieter}
\affiliation{Institute for Experimental and Applied Physics, Christian-Albrechts University in Kiel, 24118 Kiel, Germany}

\author{O. Adriani} 
\affiliation{University of Florence, Department of Physics, I-50019 Sesto Fiorentino, Florence, Italy}
\affiliation{INFN, Sezione di Florence, I-50019 Sesto Fiorentino, Florence, Italy} 

\author{G. C. Barbarino} 
\affiliation{University of Naples ``Federico II'', Department of Physics, I-80126 Naples, Italy}  
\affiliation{INFN,Sezione di Naples, I-80126 Naples, Italy}

\author{G. A. Bazilevskaya} 
\affiliation{Lebedev Physical Institute, RU-119991, Moscow, Russia}

\author{R. Bellotti}
\affiliation{University of Bari, Department of Physics, I-70126 Bari, Italy}
\affiliation{INFN, Sezione di Bari, I-70126 Bari, Italy}

\author{E. A. Bogomolov}
\affiliation{Ioffe Physical Technical Institute, RU-194021 St. Petersburg, Russia} 

\author{M. Bongi} 
\affiliation{University of Florence, Department of Physics, I-50019 Sesto Fiorentino, Florence, Italy}
\affiliation{INFN, Sezione di Florence, I-50019 Sesto Fiorentino, Florence, Italy}

\author{V. Bonvicini} 
\affiliation{INFN, Sezione di Trieste, I-34149 Trieste, Italy}

\author{A. Bruno} 
\affiliation{Heliophysics Division, NASA Goddard Space Flight Center, Greenbelt, MD, USA}
\affiliation{Department of Physics, Catholic University of America, Washington DC, USA}

\author{F. Cafagna} 
\affiliation{INFN, Sezione di Bari, I-70126 Bari, Italy}

\author{D. Campana} 
\affiliation{INFN,Sezione di Naples, I-80126 Naples, Italy} 

\author{P. Carlson} 
\affiliation{KTH, Department of Physics, and the Oskar Klein Centre for Cosmoparticle Physics,AlbaNova University Centre, SE-10691 Stockholm, Sweden}

\author{M. Casolino}
\affiliation{INFN, Sezione di Rome ``Tor Vergata'', I-00133 Rome, Italy}  
\affiliation{RIKEN, EUSO team Global Research Cluster, Wako-shi, Saitama, Japan} 

\author{G. Castellini} 
\affiliation{IFAC, I-50019 Sesto Fiorentino, Florence, Italy}

\author{C. De Santis}
\affiliation{INFN, Sezione di Rome ``Tor Vergata'', I-00133 Rome, Italy}  

\author{A. M. Galper}
\affiliation{MEPhI: National Research Nuclear University MEPhI, RU-115409, Moscow, Russia} 

\author{S. V. Koldashov} 
\altaffiliation{Deceased}
\affiliation{MEPhI: National Research Nuclear University MEPhI, RU-115409, Moscow, Russia} 

\author{S. Koldobskiy} 
\affiliation{MEPhI: National Research Nuclear University MEPhI, RU-115409, Moscow, Russia} 
\affiliation{ University of Oulu, 90570 Oulu, Finland  }

\author{A. N. Kvashnin}
 \affiliation{Lebedev Physical Institute, RU-119991, Moscow, Russia}

\author{A.A. Leonov} 
\affiliation{MEPhI: National Research Nuclear University MEPhI, RU-115409, Moscow, Russia} 

\author{V.V. Malakhov} 
\affiliation{MEPhI: National Research Nuclear University MEPhI, RU-115409, Moscow, Russia} 

\author{L. Marcelli} 
\affiliation{INFN, Sezione di Rome ``Tor Vergata'', I-00133 Rome, Italy}  

\author{M. Martucci} 
\affiliation{University of Rome ``Tor Vergata'', Department of Physics, I-00133 Rome, Italy}
\affiliation{INFN, Laboratori Nazionali di Frascati, Via Enrico Fermi 40, I-00044 Frascati, Italy}

\author{A. G. Mayorov} 
\affiliation{MEPhI: National Research Nuclear University MEPhI, RU-115409, Moscow, Russia}

\author{M. Merg\`e} 
\affiliation{University of Rome ``Tor Vergata'', Department of Physics, I-00133 Rome, Italy}
\affiliation{INFN, Sezione di Rome ``Tor Vergata'', I-00133 Rome, Italy}  

\author{E. Mocchiutti} 
\affiliation{INFN, Sezione di Trieste, I-34149 Trieste, Italy} 

\author{A. Monaco} 
\affiliation{University of Bari, Department of Physics, I-70126 Bari, Italy}
\affiliation{INFN, Sezione di Bari, I-70126 Bari, Italy}

\author{N. Mori} 
\affiliation{INFN, Sezione di Florence, I-50019 Sesto Fiorentino, Florence, Italy} 

\author{V. V. Mikhailov} 
\affiliation{MEPhI: National Research Nuclear University MEPhI, RU-115409, Moscow, Russia}

\author{G. Osteria}
\affiliation{INFN,Sezione di Naples, I-80126 Naples, Italy}

\author{B. Panico} 
\affiliation{INFN,Sezione di Naples, I-80126 Naples, Italy}

\author{P. Papini} 
\affiliation{INFN, Sezione di Florence, I-50019 Sesto Fiorentino, Florence, Italy} 

\author{M. Pearce}
\affiliation{KTH, Department of Physics, and the Oskar Klein Centre for Cosmoparticle Physics,AlbaNova University Centre, SE-10691 Stockholm, Sweden}

\author{P. Picozza} 
\affiliation{University of Rome ``Tor Vergata'', Department of Physics, I-00133 Rome, Italy}
\affiliation{INFN, Sezione di Rome ``Tor Vergata'', I-00133 Rome, Italy}

\author{M. Ricci}
\affiliation{INFN, Laboratori Nazionali di Frascati, Via Enrico Fermi 40, I-00044 Frascati, Italy}

\author{S. B. Ricciarini}
\affiliation{INFN, Sezione di Florence, I-50019 Sesto Fiorentino, Florence, Italy} 
\affiliation{IFAC, I-50019 Sesto Fiorentino, Florence, Italy}

\author{M. Simon}
\altaffiliation{Deceased}
\affiliation{Universitat Siegen, Department of Physics, D-57068 Siegen, Germany}

\author{A. Sotgiu}
\affiliation{INFN, Sezione di Rome ``Tor Vergata'', I-00133 Rome, Italy}

\author{R. Sparvoli}
\affiliation{University of Rome ``Tor Vergata'', Department of Physics, I-00133 Rome, Italy}
\affiliation{INFN, Sezione di Rome ``Tor Vergata'', I-00133 Rome, Italy} 

\author{P. Spillantini}
\affiliation{MEPhI: National Research Nuclear University MEPhI, RU-115409, Moscow, Russia} 
\affiliation{Istituto Nazionale di Astrofisica, Fosso del cavaliere 100, 00133 Roma, Italy} 

\author{Y. I. Stozhkov} 
 \affiliation{Lebedev Physical Institute, RU-119991, Moscow, Russia} 

\author{A. Vacchi}
\affiliation{INFN, Sezione di Trieste, I-34149 Trieste, Italy} 
\affiliation{University of Udine, Department of Mathematics, Computer Science and Physics Via delle Scienze, 206, Udine, Italy}

\author{E. Vannuccini}
\affiliation{INFN, Sezione di Florence, I-50019 Sesto Fiorentino, Florence, Italy} 

\author{G.I. Vasilyev} 
\affiliation{Ioffe Physical Technical Institute, RU-194021 St. Petersburg, Russia} 

\author{S. A. Voronov} 
\affiliation{MEPhI: National Research Nuclear University MEPhI, RU-115409, Moscow, Russia} 

\author{Y. T. Yurkin} 
\affiliation{MEPhI: National Research Nuclear University MEPhI, RU-115409, Moscow, Russia} 

\author{G. Zampa}
\affiliation{INFN, Sezione di Trieste, I-34149 Trieste, Italy} 

\author{N. Zampa}
\affiliation{INFN, Sezione di Trieste, I-34149 Trieste, Italy}

\begin{abstract}

Time-dependent energy spectra of galactic cosmic rays (GCRs) carry fundamental information regarding their origin and propagation. When observed at the Earth, these spectra are significantly affected by the solar wind and the embedded solar magnetic field that permeates the heliosphere, changing significantly over an 11-year solar cycle. Energy spectra of GCRs measured during different epochs of solar activity provide crucial information for a thorough understanding of solar and heliospheric phenomena. The PAMELA experiment had collected data for almost ten years (15$^{th}$ June 2006 - 23$^{rd}$ January 2016), including the minimum phase of solar cycle 23 and the maximum phase of solar cycle 24. In this paper, we present new spectra for helium nuclei measured by the PAMELA instrument from January 2010 to September 2014 over a three Carrington rotation time basis. These data are compared to the PAMELA spectra measured during the previous solar minimum providing a picture of the time dependence of the helium nuclei fluxes over a nearly full solar cycle. 
Time and rigidity dependencies are observed in the proton-to-helium flux ratios.
The force-field approximation of the solar modulation was used to relate these dependencies to the shapes of the local interstellar proton and helium-nuclei spectra.

\end{abstract}


\section{Introduction} \label{sec:intro}

Since the end of the last century, there have been a flurry of new measurements of the energy spectra and composition of the cosmic radiation with significant improvement in the statistical precision and reduction in the systematic uncertainties (for a review see \citet{Boezio_2020}). These measurements have provided new insights and breakthroughs in the investigation of the origin and propagation of galactic cosmic rays (GCRs), e.g. \citet{Blasi_2014,AmBl_2018}. Particularly significant are the measurements on protons and helium nuclei (e.g. \citet{Adriani_2011,Aguilar_2014,An_2019,CALET_2019}), the most abundant components of GCRs. However, the near totality of these measurements were obtained deep inside the heliosphere where the influence of the solar wind is especially important.

The solar wind is a plasma of ionized gas emitted by the Sun corona. 
The solar wind, whose existence was fully realized by Parker in 1958 \citep{Parker_1958}, expands at supersonic speed into space creating the heliosphere, a region of space over which our Sun influence dominates. 
Since the solar wind is coupled with the Sun corona, it carries the solar magnetic field   present in the corona out in the solar system creating the heliospheric magnetic field (HMF).
It has long been known that Solar activity has an 11-year periodicity \citep{Usoskin_2017} over which the solar wind pattern and intensity of the HMF vary significantly. During a period of minimum activity, the Sun’s global magnetic field has its simplest form, while it tends to assume a chaotic structure near maximum activity. Additionally, the solar magnetic field undergoes a polarity reversal during solar maximum resulting in a 22-year cycle for the polarity of the HMF. 

The energy spectra of the cosmic rays as measured at Earth is affected by their interaction with the turbulent solar wind and the embedded magnetic field characterizing the heliosphere. When they arrive at Earth the characteristics of the heliosphere are imprinted in their energy spectra (e.g. \citet{Potgieter_2013, Heber_2013}. Therefore, a precise understanding of the transport of GCRs in the heliosphere is required to fully exploit the precise information provided by the experimental measurements
(e.g. \citet{Potgieter:2013cwj,Potgieter:2015jxa}). Conversely, precise measurements of the cosmic-ray energy spectra down to fractions of GeV and their time dependence over a solar cycle provide unique insights on the fundamental properties of the solar wind and magnetic field turbulence in the heliosphere, the modulation of GCRs and the characteristics of solar activity.   

The PAMELA satellite-born experiment was launched from the baikonur cosmodrome in Kazakhstan on June 15th, 2006.  Then, for nearly a solar cycle, 
from the 23rd solar minimum through the maximum of
solar cycle 24, PAMELA had been making high-precision measurements of the charged component of the cosmic radiation. 
The PAMELA collaboration already published several papers on GCR solar modulation: protons \citep{Adriani_2013,Martucci_2018}, electrons and positrons \citep{Adriani_el_2015,Adriani_elpos_2016} and, most recently, the time-dependent helium spectra during the 23rd solar minimum (July 2006 - December 2009) \citep{Marcelli_2020}. In this paper, the measurement of the helium nuclei component is extended up to the end of the 24th solar maximum (September 2014). 
The new energy spectra were evaluated on a three Carrington rotations time ($\simeq 81$ days) basis from 2010 January to 2014 September; from Carrington numbers 2092 to 2154 according to the official numbering. No isotopic separation was done in this analysis, the fluxes are the sum of $^{3}$He and $^{4}$He components. 

These fluxes are combined with the previous published data to present the time dependence of the helium nuclei spectrum over a nearly complete solar cycle. 

 Additionally, for the same time period the 
proton-to-helium flux ratios are presented as a function of time and rigidity to highlight dependencies possibly due to the different particle masses and shapes of the local interstellar spectra \citep{Corti_2019,Tomassetti_2018,Ngobeni:2020quz}. 
Finally, a simplified approach to solar modulation, the force field approximation \citep{Gleeson_Axford_1968}, is used to relate these dependencies to the shapes of the local interstellar proton and helium-nuclei spectra.

\section{Instrument and data analysis} \label{sec:data_analysis}

After its launch, the PAMELA experiment had been almost continuously taking data until January 2016. The experiment was located on board the Resurs-DK1 Russian satellite placed by a Soyuz rocket at a highly inclined (70$^\circ$) elliptical orbit between 350 and 600 km height, changed into a circular one of 580 km in September 2010. The satellite quasi-polar orbit allowed the PAMELA instrument to sample low cutoff-rigidity orbital regions for a considerable amount of time, making it suitable for low-energy particle studies. The apparatus consisted of a combination of detectors that provided information for particle identification and precise rigidity ($R$) measurements. These detectors were: a Time-of-Flight system, a magnetic spectrometer, an anti-coincidence system, an electromagnetic imaging calorimeter, a shower tail catcher scintillator, and a neutron detector. Detailed information about the instrument and its performances can be found in \citet{Picozza_2007,Adriani_PhysRep_2014,Adriani_2017}. 
\\

The statistics of selected events was found to decrease over time. This effect was mainly due to the sudden, random failure of a few front-end chips in the tracking system and it became particularly significant after 2009. Therefore, in this analysis the helium fluxes were evaluated on a three Carrington rotations time basis.  
In the time period covered in this analysis the solar activity was at its maximum and characterized by many solar events. Most of these events produced high-energy particles capable to reach Earth and, consequently, the PAMELA detector. Similarly to the approach adopted in the analysis of the time dependence of GCR protons \citep{Martucci_2018}, the time periods corresponding to these solar events, according to the measurements of low-energy ($>60$ MeV) proton channel of GOES-15\footnote{\url{https://umbra.nascom.nasa.gov/sdb/goes/particle/}}, were not included in this work.
\\
The analysis procedure used in this work was identical to the one used to determine the time dependence of the helium nuclei fluxes over the solar minimum period presented and discussed in \citet{Marcelli_2020}. The absolute helium nuclei fluxes $\Phi(K)$ in kinetic energy ($K$) were obtained as follows: 
\begin{equation}
    \Phi(K)=\frac{N(K)}{G(K) \times LT \times \epsilon(K)\times \Delta K}
\end{equation}
where $N(K)$ is the unfolded count distribution of selected events, $\epsilon(K)$ the product of the single selections efficiencies, $G(K)$ the geometrical factor, $LT$ the live-time and $\Delta K$ the width of the energy interval. The total selection efficiency was $\sim22\%$ at the beginning of 2011, decreasing to $\sim12\%$ towards the end of 2014. This was mainly driven by the aforementioned condition of the tracking system. The geometrical factor for selected helium nuclei above 2 GV is 17.5 cm$^2$\ sr. 

No isotopic separation (possible only up to $\approx 1.4$ GeV/n \citep{Adriani_2016})
was performed in this analysis. 
For the conversion from rigidity to kinetic energy all helium nuclei events were treated as $^{4}$He. 

\section{Results} \label{sec:results}

\begin{figure}
\centering
\includegraphics[width=.95\textwidth]{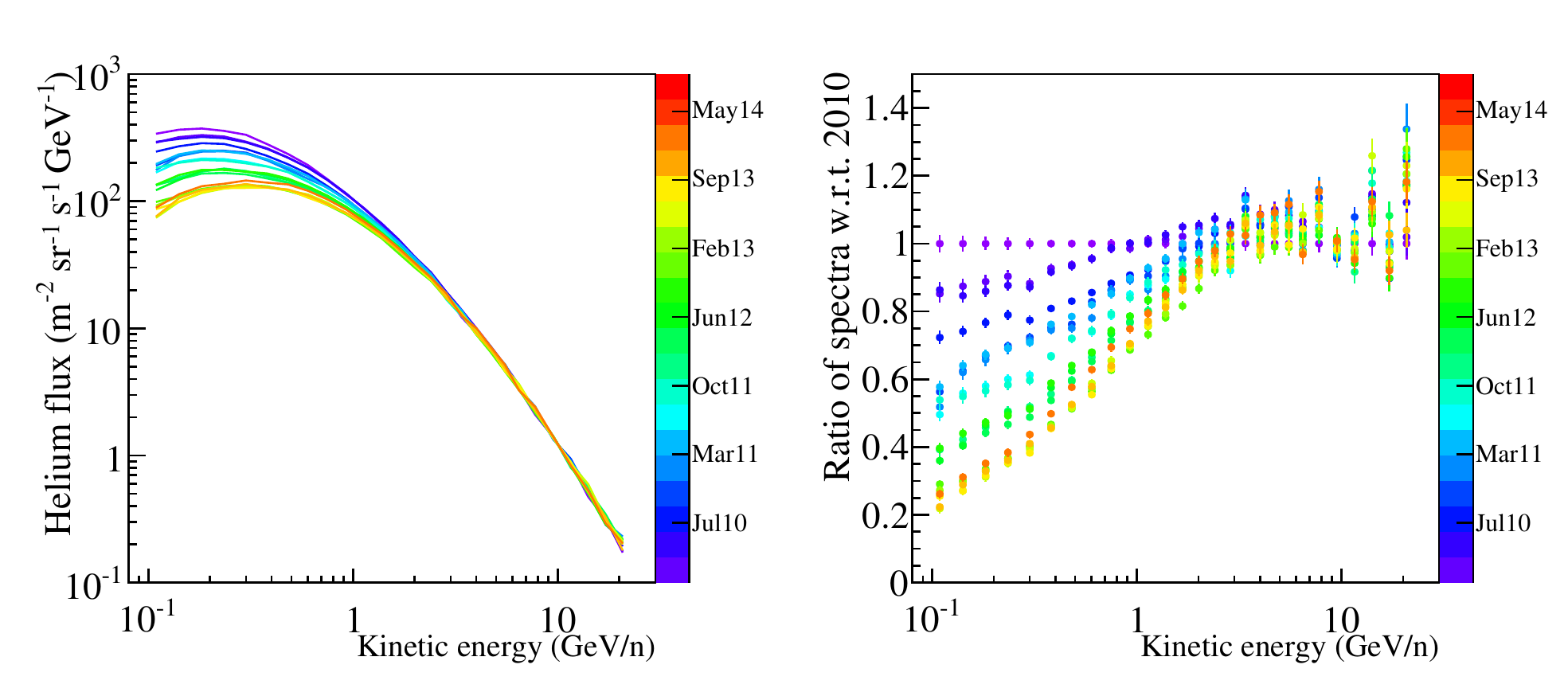}
\caption{Left panel: the evolution of the helium energy spectrum as intensities approached the period of maximum solar activity, from January 2010 (violet), to September 2014 (red). Right panel: the ratio of the measured spectra with respect to the spectrum of January 2010. The color code is the same as the right panel.}
\label{fig:rainbow_1}
\end{figure}

The resulting energy spectra are presented in Figure \ref{fig:rainbow_1}, left panel, that shows the time evolution of measured differential helium fluxes as a function of kinetic energy, from January 2010 (violet curve) to September 2014 (red curve). The right panel shows the helium flux ratio with respect to the flux measured in January 2010. The effect of solar modulation is clearly visible in the energy region below few GeV/n where it causes the flux to decrease significantly and subsequently modifies the spectral shape with increasing solar activity.

The flux intensity measured at the energy interval $95-123$\ MeV/n dropped by about $70\%$ from January 2010 to September 2014, while the flux intensity at $337-427$\ MeV/n decreased  about $50\%$ during the same time interval. At energies above $\sim 15$ GeV/n the solar modulation effect is assumed negligible with respect to the experimental uncertainties.

\begin{figure}
    \centering
    \includegraphics[width=.95\textwidth]{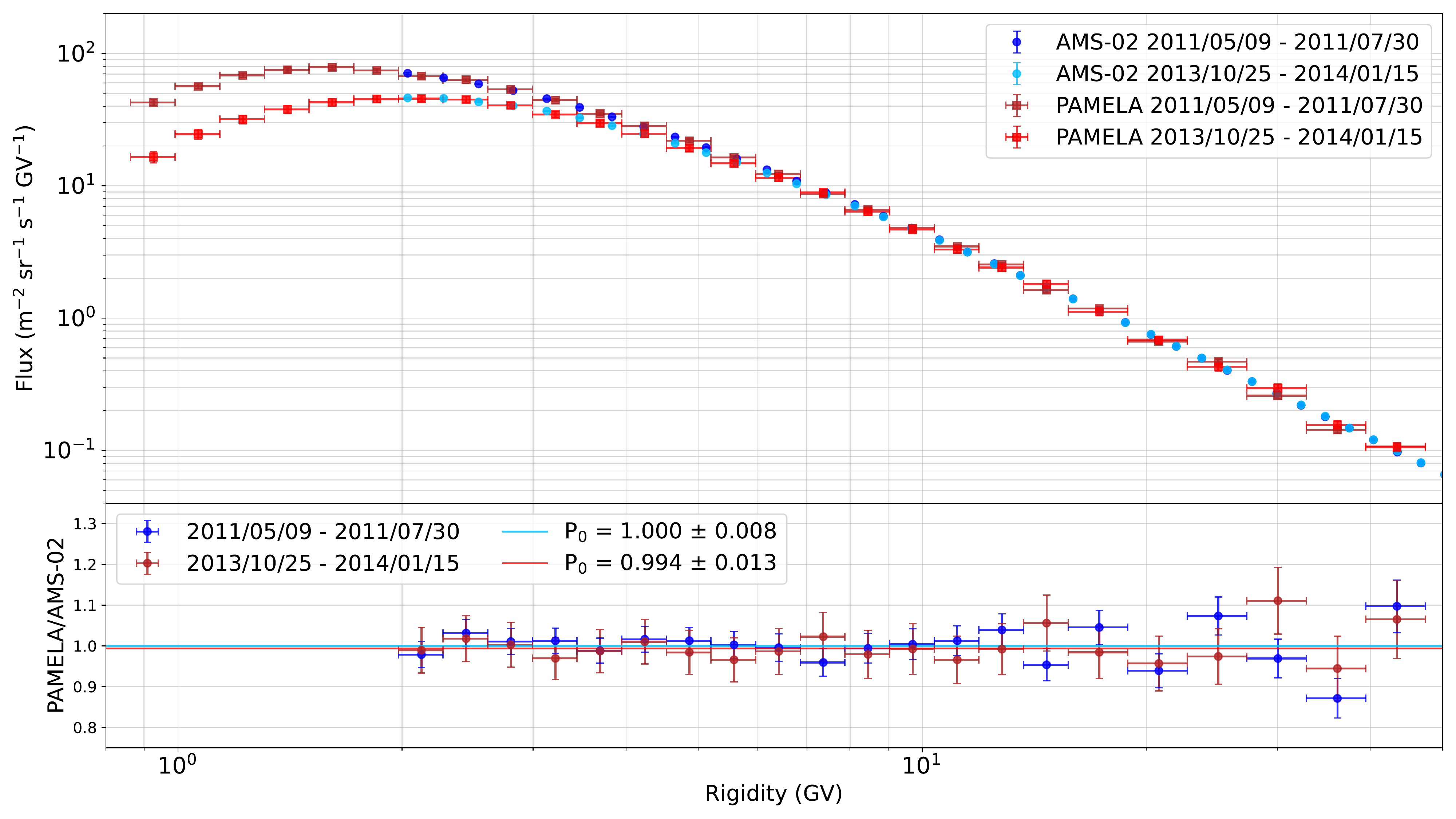}
    \caption{Top: PAMELA helium spectra measured in 2011 and late 2013 compared with the corresponding AMS-02 measurements \citep{Aguilar_2018}, as indicated in the legend. The AMS-02 fluxes were averaged over 3 Carrington Rotations. Bottom: the ratios between PAMELA and AMS-02 fluxes. A constant fit was performed on the ratios and the results ($P_0$) are shown in the legend.}
    \label{fig:pam_ams_comp}
\end{figure}
Figure \ref{fig:pam_ams_comp} shows a comparison between two PAMELA and AMS-02 helium fluxes measured during Carrington Rotation numbers 2110-2112 and 2143-2145. The published AMS-02 fluxes \citep{Aguilar_2018} were averaged over 3 Carrington Rotations to match the PAMELA time periods.
An excellent agreement between the two sets of measurements can be noticed in the overlapping rigidity region as shown in the bottom panel by the constant fits on the ratios between PAMELA and AMS-02 measured fluxes.

\begin{figure}
\centering
\includegraphics[width=.95\textwidth]{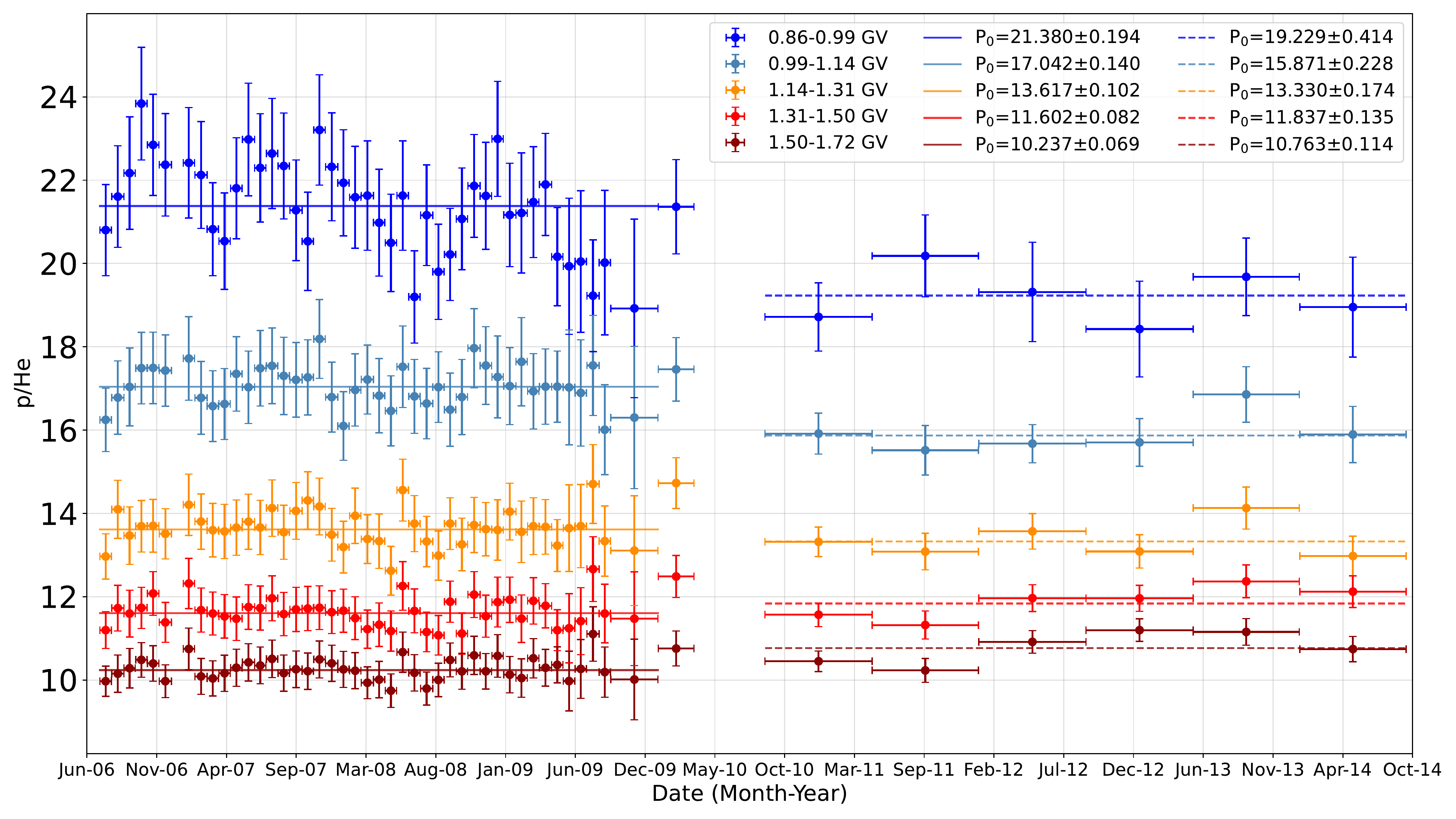}
\caption{Time profiles of proton-to-helium ratio for the five rigidity intervals specified in the legend. The error bars are the quadratic sum of statistical and systematic errors. After 2009 the data points cover 9 Carrington Rotation time periods. The lines and $P_0$ parameters resulting from the fit of a constant in each rigidity bin in the solar minimum (solid line) and maximum (dashed line) periods are also shown.}
\label{fig:phe_ratio}
\end{figure}

The time dependencies of protons and helium nuclei were analyzed by measuring the proton-to-helium flux ratio as a function of time and rigidity.   
Figure \ref{fig:phe_ratio} shows this ratio for five rigidity intervals for increased solar activity, this work, and for the preceding solar minimum period \citep{Marcelli_2020}.
Since the quantity measured by the magnetic spectrometer is rigidity, this approach allows a more precise estimation of the ratios considering that systematic uncertainties, related to the same instrumental effects, cancel out. The residual systematic uncertainty includes only the errors due to the efficiency estimation. The error bars in Figure \ref{fig:phe_ratio} are the quadratic sum of the statistical errors and this residual systematic error. To reduce the statistical fluctuation for the data points after 2009, a weighted average over nine months was performed, while the data points relative to the solar minimum
are shown with the original time basis, described in \citet{Marcelli_2020}. 
Each rigidity interval of the two data sets (until 2009 and from 2010) were fit with a constant ($P_0$) function, whose fitted lines and values are shown in Figure \ref{fig:phe_ratio}.
There is a clear evidence of a decrease in time for the lowest rigidity intervals 
from minimum to maximum solar activity period. On the contrary, the higher rigidity intervals 
show hints of the opposite behavior as the ratios increase from  minimum to  maximum solar activity.

\begin{figure}
\centering
\includegraphics[width=.95\textwidth]{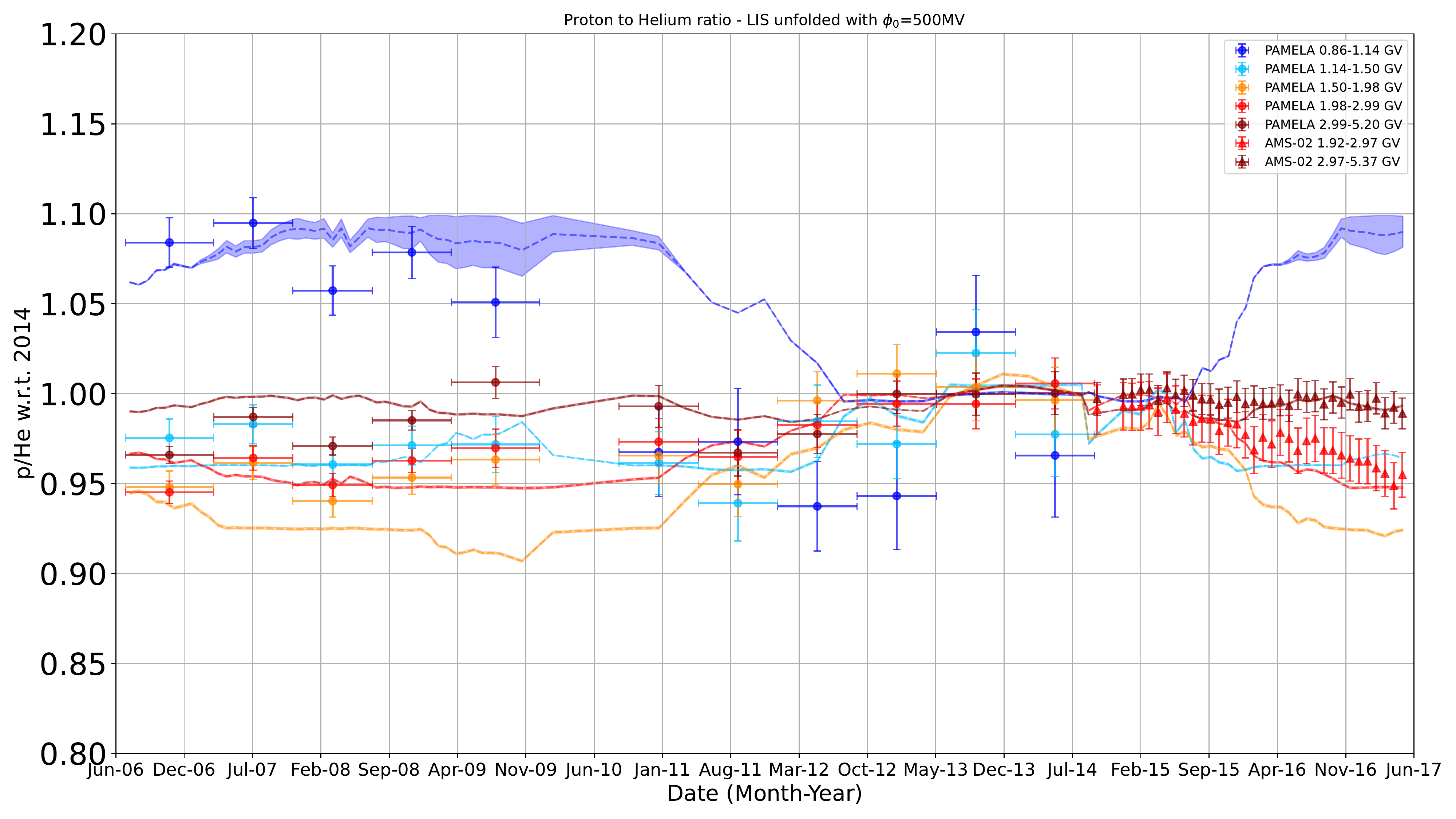}
\caption{Time profile of the proton-to-helium ratio at the  rigidity intervals listed in the legend, normalized to the mean value of the ratios in the solar maximum period from May 2013 to September 2014. The error bars are the quadratic sum of statistical and systematic errors. A weighted average over nine month
(with the exception of the last point before 2010 which is a weighted average over ten months)
was performed to reduce statistical fluctuation and the effect of the short-term cyclic variation in the solar minimum period. The AMS-02 data \citep{Aguilar_2018}, normalized over the same solar maximum period, are shown for the period after September 2014. The dashed lines are the proton-to-helium ratios and their corresponding uncertainties, colored bands, derived using the force-field approximation for solar modulation as described in Sec. \ref{sec:interpretation}.}
\label{fig:phe_norm}
\end{figure}

Finally, Figure \ref{fig:phe_norm} shows the PAMELA proton over helium ratio for a wider rigidity range
and normalized to the mean value of the ratio in the period from May 2013 to September 2014, when the solar activity reached its maximum. To reduce the effect of the short-term cyclic variation in the solar minimum period, mainly visible in the lowest rigidity interval in Figure \ref{fig:phe_ratio}, a weighted average over a nine-month time basis was performed. The last point before 2010 corresponds to a weighted average over ten months. The AMS-02 data \citep{Aguilar_2018}, normalized over the same solar maximum period,
are also shown for the period after September 2014, with the rigidity bins combined to better match those of PAMELA data. A clear time dependence is observed for the lowest rigidity bin (blue), which decreases from minimum to maximum solar activity period. Conversely, above about $1.5$ GV (red and yellow) the proton-to-helium ratio shows an increase from 2006 to 2014 followed by a comparable decrease seen in the AMS-02 data. These time and rigidity dependencies can be related to effects caused by the solar modulation of protons and helium because of the difference in their mass-to-charge ratio and in the shape of their respective LIS. For an illustration of these modulation effects, see \citet{Ngobeni:2020quz}).

A detailed theoretical modelling of the proton-to-helium flux ratio will be the topic of a future publication. In this work a first analysis of the relevance of the LIS shapes for the features of this ratio was conducted   
using the force-field approximation for solar modulation.

\section{Data interpretation and discussion} \label{sec:interpretation}

Assuming the force-field approximation, appropriate for GCR with kinetic energies above about 200 MeV/n, of the spherically symmetric model for solar modulation suggested by \citet{Gleeson_Axford_1968}, the differential GCR intensity $J(r,E,t)$ at a given distance $r$ from the Sun, total energy $E$ and time $t$, is related with the time-independent interstellar GCR intensity $J(\infty,E)$ through the equation:
\begin{equation}
    \label{eq:ffe}
    J(r,E,t) = \frac{E^2-E_0^2}{(E+\Phi)^2-E_0^2}J(\infty,E+\Phi(t))
\end{equation}

where $E_0$ is the rest energy (mass) of the particle and $\Phi=|Z|e\phi$ a parameter that can be interpreted as the energy loss experienced by the cosmic-ray particle when approaching the Earth from infinity. Therefore, the time dependence of the GCR fluxes due to solar modulation is reproduced by the time dependence of the solar modulation parameter $\phi$. 
Consequently, if $\phi$ is known, the LIS can be extrapolated from the modulated spectrum using equation~\ref{eq:ffe} .

\begin{figure}
    \centering
    \includegraphics[width=.95\textwidth]{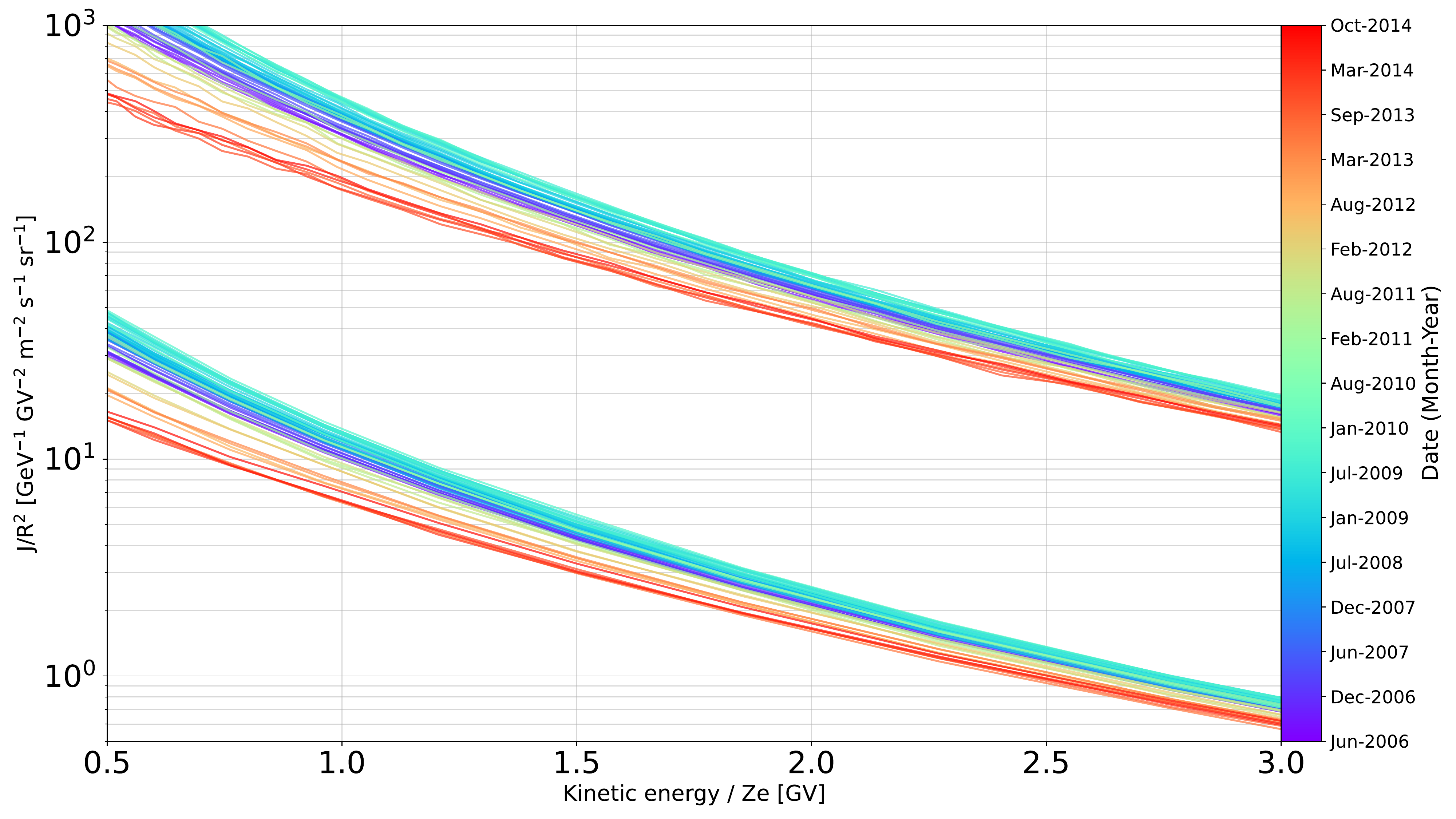}
     \caption{Proton (top set of lines) and helium nuclei (lower lines) fluxes ($J$) measured by the PAMELA experiment from June 2006 (violet) to September 2014 (red) divided by rigidity squared versus kinetic energy divided by the charge (Ze). The displacements along the abscissa are similar between the two species and they are attributed to the varying with time of the solar modulation parameter.}
    \label{fig:jr2}
 \end{figure}

In this work the solar modulation parameters were obtained with the following procedure  (for more details see \citet{Marcelli_2021_PhD}). 
Following \citet{Gleeson_Axford_1968}, two sets of curves,  shown in Figure~\ref{fig:jr2},  were obtained by plotting the measured proton and helium nuclei fluxes divided by rigidity squared, i.e. $J/R^2$, as a function of the kinetic energy divided by the particle charge, i.e. $K/Ze$. As can be seen in Figure~\ref{fig:jr2}, above 0.5 GV, these curves have similar shape but displaced along the abscissa. 
These displacements represent the time-dependent change, $\Delta \phi$, in the solar modulation parameter ($\phi = \phi_0 + \Delta \phi$). From these curves, the $\Delta \phi$ were obtained for both particle species
and, as expected \citep{Gleeson_Axford_1968}, inside the experimental uncertainties were found identical and comparable with the variations of the solar modulation parameter determined by \citet{Koldobskiy:2019olx} using neutron monitor, AMS-02 and PAMELA data.
Subsequently, a set of LIS, one for each measured modulated spectrum, was estimated assuming a $\phi_0$ value of 300 MV and, then, merged into a single spectrum with a weighted average procedure. 
The flux values of this resulting LIS were combined with the Voyager 1 data \citep{Stone_2013, Cummings_2016} at lower energies. Then, the value of $\phi_0$ was increased at steps of 10 MV and for each step a new combined LIS was obtained. The LIS that had the smoothest spectrum data\footnote{The smoothness of each spectrum was obtained by dividing the standard deviation of the differences between consecutive flux values by the mean of these difference. Then the smoothest spectrum was the one with the minimum value for the smoothness.} provided the 
best value for $\phi_0$, which was found to be 500 MV both for protons and helium nuclei. 
Consequently, the best LIS for the two-particle species were also obtained.
Finally, these LIS were modulated with the estimated modulation parameters for the period June 2006-June 2017 
and the proton-to-helium flux ratios of the resulting modulated fluxes were calculated. For the period October 2014-June 2017 the solar modulation parameters estimated by \citet{Koldobskiy:2019olx} were used.
These ratios are shown in figure \ref{fig:phe_norm} as dashed curves along with the propagated uncertainties (sum of statistical and systematic errors) of the measured fluxes shown as colored bands. 

Considering the significant approximation of the force-field approach, it is worth noticing that the calculated proton-to-helium flux ratios qualitatively reproduce the time and rigidity dependencies observed with the experimental data of both PAMELA (until September 2014) and AMS-02 (after September 2014). 

Since the force-field approximation assumes the same modulation parameter for different particle species, this result would indicate that,  
 in the rigidity range of these measurements, the observed time variation of the proton-to-helium flux ratios are dominated by the shapes of the proton and helium nuclei LIS, while the dependence of the diffusion tensor of the heliospheric transport equation on the particle mass-to-charge ratio would appear to play an increasing role at lower rigidities.

\section{Conclusions} 

The PAMELA experiment observed GCR data for nearly a complete solar cycle from the minimum phase of solar cycle 23 to the maximum phase of solar cycle 24. In this work, we have presented new spectra for helium nuclei measured by the PAMELA instrument from January 2010 until September 2014 integrating the previously published data. These  measurements allow a detailed study of the propagation of cosmic rays inside the heliosphere. 
Comparing the helium nuclei fluxes to the proton fluxes, time and rigidity dependencies are clearly observed. A quantitative study, based on state-of-the-art models (e.g. \citet{Ngobeni:2020quz}), of these dependencies is underway and will be presented in a future publication. 
However, a simplified approach based on the force field approximation of solar modulation was able to 
relate these dependencies to the shapes of the local interstellar proton and helium-nuclei spectra. 

The results discussed in this paper will be available at the Cosmic Ray Data Base of the ASI Space Science Data Center (http://tools.asdc.asi.it/CosmicRays/chargedCosmicRays.jsp).

\begin{acknowledgments}

We acknowledge partial financial support from The Italian Space Agency (ASI). We also acknowledge support from Deutsches Zentrum fur Luft- und Raumfahrt (DLR), The Swedish National Space Board, The Swedish Research Council, The Russian Space Agency (Roscosmos), Russian Science Foundation No-20-72-10170, RFBR and Russian ministry of science.
R.M. acknowledges partial financial support from the INFN Grant "giovani", project ASMDM.

\end{acknowledgments}

\newpage

\bibliography{Bibtex_solmod}{}
\bibliographystyle{aasjournal}

\end{document}